\documentclass[aps,prl,twocolumn,showpacs,superscriptaddress,
preprintnumbers,amsmath,amssymb]{revtex4}

\usepackage{graphicx}
\usepackage{dcolumn}
\usepackage{bm}
\usepackage{ifthen}
\usepackage{booktabs}

\begin{document}

\title{Origin of ferroelectricity in
high-$T_c$ magnetic ferroelectric CuO}

\author{Guangxi Jin}
\affiliation{Key Laboratory of Quantum Information, University of
Science and Technology of China, Hefei, 230026, People's Republic of
China}

\author{Kun Cao}
\affiliation{Key Laboratory of Quantum Information, University of
Science and Technology of China, Hefei, 230026, People's Republic of
China}

\author{Guang-Can Guo}
\affiliation{Key Laboratory of Quantum Information, University of
Science and Technology of China, Hefei, 230026, People's Republic of
China}

\author{Lixin He}
\email{helx@ustc.edu.cn}
\affiliation{Key Laboratory of Quantum Information, University of
Science and Technology of China, Hefei, 230026, People's Republic of
China}

\date{\today}

\begin{abstract}

Cupric oxide is a unique magnetic ferroelectric material with a
transition temperature significantly higher than the boiling point
of liquid nitrogen. However, the mechanism of high-T$_c$
multiferroicity in CuO remains puzzling. In this paper, we clarify
the mechanism of high-T$_c$ multiferroicity in CuO, using combined
first-principles calculations and an effective Hamiltonian model. We
find that CuO contains two magnetic sublattices, with strong
intrasublattice interactions and weakly frustrated intersublattice
interactions,
which may represent one of the main reasons for the high ordering temperature of the compound.
The weak spin frustration leads to incommensurate spin excitations that dramatically
enhance the entropy of the mutliferroic phase and eventually stabilize that phase in CuO.
\end{abstract}



\maketitle

``Magnetic ferroelectric'' materials, in which ferroelectricity is induced
by magnetic ordering, have attracted intense interest
\cite{fiebig05,cheong07}. The strong magnetoelectric (ME) coupling
in these materials opens a new path to the design of multifunctional
devices that allow the control of charges by the application of
magnetic fields or spins through applied voltages. However, nearly
all current magnetic ferroelectric materials are strongly frustrated magnets
\cite{cheong07}, with very low ordering temperatures ($\sim$ 30 - 40
K), several times lower than the temperatures expected from the
strengths of their spin interactions. Low critical temperature is
one of the major factors that limit the application of these
important materials. Therefore, a new mechanism that allows
high-temperature magnetic ferroelectricity is highly desirable.

\begin{figure}
\centering
\includegraphics[width=2.8in]{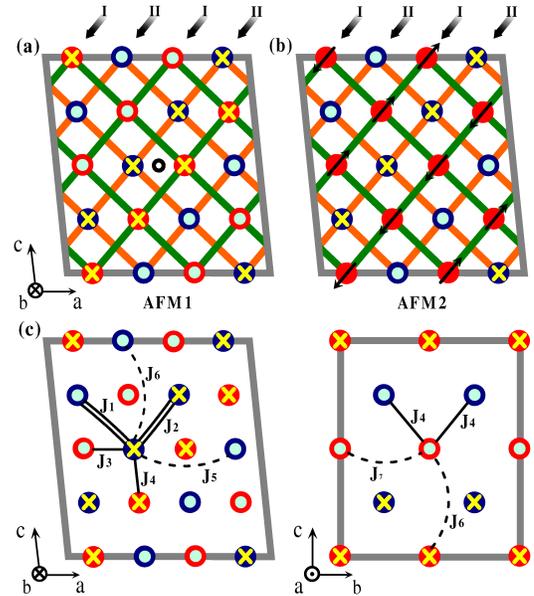}
\caption{Schematic sketch of the magnetic structures of (a) the collinear AFM1
  phase, and (b) the noncollinear AFM2 phase.  The black arrows, yellow crosses, and
  blue circles denote the spin directions associated with Cu ions.
  The black circle in (a) indicates an inversion center.
 (c) A sketch of superexchange interactions $J_1$ to $J_7$.
  The single lines, double lines, and dashed lines represent the three types of
  exchange interactions between Cu ions.}
\label{fig:spin}
\end{figure}

Recently, CuO has been found to be multiferroic at $T_c$=230 K,
which is much higher than the critical temperatures of any other
magnetic ferroelectric materials \cite{kimura08}. CuO undergoes two
successive magnetic phase transitions when it is cooled from room
temperature to a temperature near zero. Neutron scattering
experiments \cite{yang89} have shown that at temperatures below
$T_{N1}$=213 K, the spin structure is collinear antiferromagnetic
(AFM1) [see Fig. \ref{fig:spin}(a)]. Between $T_{N1}$ and
$T_{N2}$=230 K, the spin structure becomes non-collinear and
slightly incommensurate (AFM2) [see Fig. \ref{fig:spin}(b)], with a
modulation vector of ${\bf Q}$= (0.006, 0, 0.017).
An electric polarization of 160 $\mu$C m$^{-2}$ along the b axis also
develops in the AFM2 phase.
Elucidation of this unusual high-temperature multiferroic behavior may provide
useful information in the search for novel room-temperature
magnetic ferroelectric materials. However, the mechanism that
stabilizes the multiferroicity in CuO is still not understood, and
it remains the subject of significant debate
\cite{giovannetti11,toledano11}.
The phase diagram of CuO has been studied via a phenomenological approach by
Tol\'{e}dano et al. \cite{toledano11}, which based on symmetry considerations
only, could not reveal the microscopic mechanism that stabilizes the multiferroic
phase in CuO. Giovannetti et al. performed Monte Carlo simulations on an effective
Hamiltonian model of CuO \cite{giovannetti11}. However, in the simulation the spins
were artificially constrained into only four possible directions, which dramatically
changes the free energy of the system and misses important physics in CuO.

In this work, we clarify the mechanism of high-T$_c$ multiferroicity
in CuO using combined first-principles calculations and simulations
based on an effective Hamiltonian model. We find that CuO contains
two magnetic sublattices, with strong intrasublattice interactions
and weakly frustrated intersublattice interactions,
 which might represent one of the main reasons that the compound exhibits a high ordering temperature.
The weak spin frustration leads to
incommensurate spin excitations that dramatically enhance the
entropy of the AFM2 phase and eventually stabilize the mutliferroic
phase. This mechanism is novel and differs from previously proposed
mechanisms \cite{giovannetti11,toledano11}.
This work suggests that high- $T_c$ magnetic ferroelectric materials can
be sought in weakly frustrated magnets similar to CuO.

The crystal structure of CuO is monoclinic and contains four
chemical units per unit cell. The AFM1 spin structure is composed
of two antiferromagnetic (AFM) spin sublattices, shown in two
different colors in Fig. \ref{fig:spin} (a), in which Cu ions have
the same $b$ values in each sublattice. The spin chains along the
[10$\bar{1}$] direction are antiferromagnetic, and are labeled chain
I and chain II for the two sublattices, whereas the chains along the
[101] direction are ferromagnetic. In the AFM1 phase, all spins are
aligned in the $b$ direction, whereas in the AFM2 phase, chain II
rotates perpendicularly to chain I.

We perform ab initio calculations on CuO, with non-collinear
spin-polarized local density approximation (LSDA) implemented in the
Vienna ab initio simulations package (VASP)
\cite{kresse93,kresse96}. The on-site Coulomb interactions $U$=7.5
eV are included for Cu ions in a rotationally invariant scheme
\cite{anisimov91}. The
spin-orbit coupling is considered in the calculation unless
otherwise noticed. To accommodate the spin structures, we use a
2$\times$1$\times$2 CuO supercell that contains 32 atoms. For the
AFM2 structure, we neglect the small incommensurate component of the
spin structure [i.e., we set ${\bf Q}$= (0, 0, 0)], and rotate the
spin directions of chain II 90$^\circ$, so that it lies in the $ac$
plane. The spins form cycloidal spirals along both the $a$ and $c$
axes in the AFM2 phase.  The incommensurate component of the magnetic
modulation vector  ${\bf Q}$ is extremely small, and it should not
affect the calculated electric polarization because $P \propto S_i
\times S_j$ \cite{mostovoy06}. More details of the first-principles
calculations can be found in Ref.\cite{jin10}.

We first determine the crystal structure of CuO under the AFM1 spin
configuration. The room temperature crystal structure is monoclinic
and of space group C2/c, with inversion symmetry. However, the
magnetic structure of the AFM1 phase only has P2$_1$/c symmetry.
Therefore, after relaxation, the crystal structure is also reduced
to P2$_1$/c symmetry because of ``exchange striction'' effects
\cite{wang08}. Further analyses \cite{wang08} show that the Cu and O
ions deviate from their high symmetry sites by approximately
10$^{-3}$ \AA. This distorted structure preserves the inversion
symmetry; therefore it has no net polarization. The inversion center
is shown in Fig. \ref{fig:spin}(a). An inversion operation about the
inversion center changes spin chain I to chain II.

We then fix the spin orientations to the AFM2 configuration and relax
the crystal structure again to obtain the crystal structure of the
AFM2 phase.
It turns out that the AFM2 configuration is a local minima of the total energy.
The results change little if the spins are let free to rotate after the initial
spin configuration is set to the AFM2 phase.
The calculated total energy of the AFM2 phase is greater
than that of the AFM1 phase by approximately 0.33 meV per atom, which
is consistent with experiments that have shown the AFM2 phase to
appear at a higher temperature than that at which the AFM1 phase
appears\cite{yang89, brown91, ain92}. Before structural relaxation,
the total energy of the AFM2 phase is approximately 0.08 meV per atom
higher than that of the AFM1 phase. This energy difference is
primarily attributed to the spin anisotropy energy, which can be
observed when the spin-orbit coupling is turned off, which reduced
the energy difference between the two phases to 0.03 meV per atom.
The remaining difference of 0.25 meV per atom exists because the
ionic distortion in the AFM1 phase is substantially larger than that
in the AFM2 phase, as shown below.

The AFM2 spin structure exhibits P2$_1$ space group symmetry, to
which the rotation of spin chain II does not give inversion
symmetry, as can be seen in Fig. \ref{fig:spin}(b). The crystal
structure is distorted by the Dzyaloshinskii-Moriya (DM) interaction
\cite{dzyaloshinskii64,moriya60,sergienko06}, which breaks the
inversion symmetry. When compared with the high symmetry structure,
all the oxygen ions are shifted in the +$b$ direction by
approximately 7$\times$10$^{-5}$ \AA, whereas all Cu ions are
shifted in the -$b$ direction by a similar amount. The ionic
distortion in the AFM2 phase is approximately two orders of magnitude
smaller than that driven by the ``exchange striction'' effects in
the AFM1 phase.

Next, we calculate the electric polarization using the Berry-phase
theory of polarization \cite{king-smith93}. The calculated total
polarization is approximately 90 $\mu$C m$^{-2}$ in the -$b$
direction, which is somewhat smaller in magnitude than the
experimental value of 160 $\mu$C m$^{-2}$ along the $b$ axis. The
agreement between the theoretical calculations and experimental
values is reasonable, given that the current functionals are not
adequate to treat the subtle correlation effects in
magnetic ferroelectric materials \cite{moskvin08}.

\begin{figure}
\centering
\includegraphics[width=2.3in]{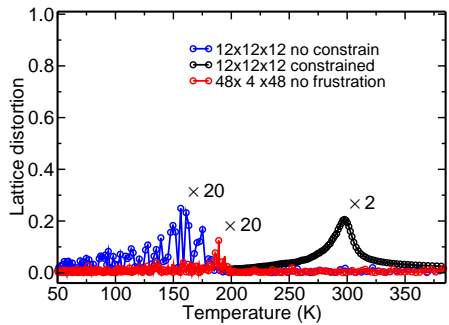}
\caption{The polar lattice distortions $u_2$ as functions of
  temperature.
The maximum value of the lattice distortion is set to 1. Blue curve:
the result of the 12$\times$12$\times$12 lattice. Red curve: the
result of the 48$\times$4$\times$48 lattice without frustrated spin
interactions. Black curve: the result of the 12$\times$12$\times$12
lattice, with the spins are constrained in 4 perpendicular
directions.
 The lattice distortion
values are amplified by factors of 20, or 2 as indicated in the
figure. } \label{fig:phase-diag1}
\end{figure}

In magnetic ferroelctric materials, the transition temperatures are
predominantly determined by the magnetic exchange interactions
\cite{kimura08}. We extract the superexchange interactions $J$s of
CuO using a Heisenberg model, $H_{\rm M}=-\sum_{ij} J_{ij} {\bf S}_i
\cdot {\bf S}_j- \sum_i (\textbf{K}\cdot\textbf{S}_i)^2$ from the
calculated total energies of the different spin configurations in
the symmetrized C2/c {\it crystal} structure with spin-orbit coupling. ${\bf K}$ is
the anisotropic energy due to spin-orbit coupling. There are seven $J$ values in
total, which are shown in Fig. 1(c). Among them, $J_1$ is the
exchange interaction between nearest-neighbor Cu atoms along
the [10$\bar{1}$] direction, $J_2$ is the interaction between nearest-neighbor
Cu atoms along the [101] direction, $J_7$ is the interaction between the 
nearest-neighbor
spins of the same sublattice along the $b$ direction, and $J_3$ and
$J_4$ are the inter-sublattice exchange interactions. The fitted
value of $J_1$= -51 meV is in good agreement with the value
determined from neutron scattering experiments, $J$=67$\pm$20
meV\cite{yang89}. The fitted value of $J_2$=8.6 meV, is only
approximately 1/6 the value of $|J_1|$, which is consistent with the
quasi-1D model \cite{shimizu03}, and the fitted value of $J_7$=9.87
meV.
The fitted inter-sublattice coupling value $J_3$=4.9 meV
and $J_4$=7 meV are weak, because the Cu-O-Cu bond angles are close
to 90$^\circ$ for these two $J$s \cite{shimizu03}. In the AFM1 and
AFM2 phases, symmetry causes additional mutual cancellation of $J_3$
and $J_4$; therefore, the energy cost of chain II rotation is low,
as was discussed in the previous paragraph. We also calculate the
next-nearest-neighbor interactions, $J_5$ and $J_6$. We find
that $J_5$ and $J_6$ are highly asymmetric. Interaction $J_5$ is
significant, with a value of -12 meV, whereas $J_6$ is only 2.1 meV.
These values agree well with those of Filippetti and Fiorentini
\cite{filippetti05}.

The intrasublattice interactions $J_1$, $J_2$ and $J_7$
essentially determine the ground state spin structure, AFM1. The
$J_5$ next-nearest-neighbor interactions further favors the antiferromagnetic spin
chain along the [10$\bar{1}$] direction, whereas, $J_6$ only adds a
small frustration to this configuration. The major competing
interactions are those of the inter-sublattice interactions, $J_3$
and $J_4$.
The weak incommensurateness of the spin spiral caused by the
frustrated exchange interactions $J_3$, $J_4$ is consistent with
that effect in this material, because the spin competition is small.
We calculate the ordering temperature of the Heisenberg model by a
Monte Carlo simulation, in which all exchange interactions are
forced to be ferromagnetic and we obtain $T_c$= 311 K
\cite{footnote2}. This temperature is only about 1.5 times greater
than the $T_c$ value of the AFM1 phase, (in RMn$_2$O$_5$
\cite{cao09}. the ratio is approximately 3 - 4.), which also
indicates that the spin frustration is weak in CuO. The lack of
strong competing interactions in this compound may explain the high
spin-ordering temperature of CuO.

To study the mutliferroic phases of CuO, we simulate the full
Hamiltonian model with spin-lattice interactions, $H=H_{\rm ph} +
H_{\rm M} + H_{\rm I}+ H_{\rm DM}$, where,
\begin{eqnarray}
H_{\rm ph}  &=& E_0 + \sum_k
 \frac{1}{2}m_1 {\omega}_1^2
u_1^{2}(k) + \frac{1}{2}m_2 {\omega}_2^2
u_2^{2}(k) \, ,\\
H_{\rm I} &=&\sum_{i,j}\sum_{k}J_{ij}'
u_1 (k)\textbf{S}_i\cdot\textbf{S}_j \, , \\
H_{\rm DM} &=& \sum_{i,l>0} D [u_2 (i) {\hat {\bf b}} \times {\bf e}_{i, i+l}]
\cdot({\textbf{S}_i}\times{\textbf{S}_{i+l}}) \, .
\label{eq:eff_h1}
\end{eqnarray}
Here $E_0$ is the energy of the high-symmetry structure without magnetic interactions.
Since the symmetry-lowering displacement is extremely small, we treat this displacement
in the AFM1 (AFM2) phase as the nonpolar(polar) phonon mode $u_1$ ($u_2$)
\cite{cao09}. $u_1(k)$ and $u_2(k)$ are the phonon modes in the kth unit cell,
and $m_1$ ($m_2$) and $\omega_1$ ($\omega_2$) are the reduced mass and
frequency of the non-polar modes (polar modes), respectively.
For simplicity, we neglect the phonon dispersion in the simulation,
which has little effect on the results\cite{cao09}. $H_{\rm I}$ is
the isotropic spin-lattice interaction caused by exchange
striction effects, $J_{ij}^{\prime} = \frac{\partial J_{ij}}{\partial u_1}$.
In CuO, only $J_3'$ and $J_4'$ are involved in
the nonpolar lattice distortion, and all other $J'$s are canceled out
by symmetry. $H_{\rm DM}$ is the DM interaction term, which
sums over the nearest neighbor spin $l$ along the $a$ and $c$ axes,
where ${\hat {\bf b}}$ is the unit vector along the $b$ (polar) axis
and ${\bf e}_{i, i+l}$ is a unit vector connecting the $i$-th and the
$i+l$-th Cu atoms. Unlike the model in Ref. \cite{giovannetti11},
our model explicitly includes the lattice degree of freedom,
especially the non-polar modes, which are missing in the model of
Ref. \cite{giovannetti11}.

For simplicity, we redefine $u_1$ ($u_2$) to be a dimensionless
parameters that takes the value of unity at the low-symmetry state
of the AFM1 (AFM2) phase, and we assign spin moments $|{\bf
S}_i|$=1.0. All parameters in the model can be obtained by fitting
to the total energies of the first-principles calculations
\cite{cao09}. We use $J_3'$=$J_4'$=0.8163 meV, which is
approximately 1.5 times greater the values obtained by
first-principles calculations \cite{footnote1}. To obtain the
correct ground state, we reduce the values to $J_3$=2.45 meV, and
$J_4$=3.5 meV, which are approximately half of the fitted values.
$m_1\omega_1^2$=0.4, $m_2\omega_2^2$=0.49$D$. $D_c$=0.8723 meV is
the critical value at which the energies of AFM1 and AFM2 phases are
degenerate.
Our main results are valid in a reasonable range of the values of
parameters for the model Hamiltonian, provided $J_3^{\prime}$ and $J_4^{\prime}$ are
larger than 0.27$J_3$ (to ensure that the ground state is in the AFM1 phase)
and $J_3$ and $J_4$ are larger than half of the fitted values.

\begin{figure}
\centering
\includegraphics[width=2.3in]{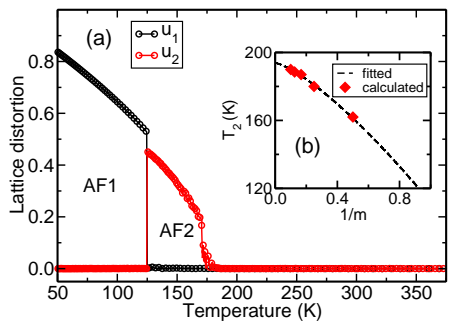}
\caption{ (a) The lattice distortions $u_1$ (black) and $u_2$ (red) as functions of
  temperature. The maximum value of the lattice distortion is set to 1.
(b) The transition temperature $T_2$ scaling with different simulation size $m$. }
\label{fig:phase-diag2}
\end{figure}

We simulate the effective Hamiltonian model in the temperature range
from 29 to 387 K, using a replica-exchange Monte Carlo method
\cite{cao09}. We perform the Monte Carlo simulations on the $n \times m
\times n$ lattices ($n$=12 - 72, $m$=2 - 12) with periodic boundary
conditions. The electric polarization $P= \langle u_2 \rangle$.
Typical simulation results for the $n$=12 lattice are shown in Fig.
\ref{fig:phase-diag1}. There is no stable AFM2 phase, even with a
large DM interaction, $D$=0.5$D_c$. If we {\it artificially}
constrain the spin to only four possible directions at 90$^\circ$,
as was done by Giovannetti at al. \cite{giovannetti11}, we obtain a
{\it weak} AFM2 phase at approximately 300 K, as shown in Fig.
\ref{fig:phase-diag1}, that is similar to the one obtained by
Giovannetti et al. \cite{giovannetti11}. The above results suggest
that the DM interaction alone cannot stabilize the AFM2 phase in CuO.

Surprisingly, if the simulation lattice size is increased to $n>$32,
a well defined AFM2 phase is obtained. The simulation results of a
48$\times$4$\times$48 lattice with a small $D$=0.066 $D_c$ are
presented in Fig. \ref{fig:phase-diag2}(a). A paraelectric (PE) to ferroelectric transition
clearly occurs near $T_2$=180 K. At $T_1$= 125 K, the polarization
suddenly drops to zero, accompanied by the appearance of the
non-polar lattice distortion, which indicates that the system enters
the AFM1 phase. This result is in excellent agreement with
experiments. Fourier analyses of the spin structures \cite{cao09}
also confirm the above results. A finite size scaling for $T_2$ with
$m$ gives $T_2$=192 K as $1/m$ approaches zero [see Fig.
\ref{fig:phase-diag2}(b)], which slightly underestimates the
experimental value of $T_2$=230 K \cite{kimura08}.

\begin{figure}
\centering
\includegraphics[width=2.5in]{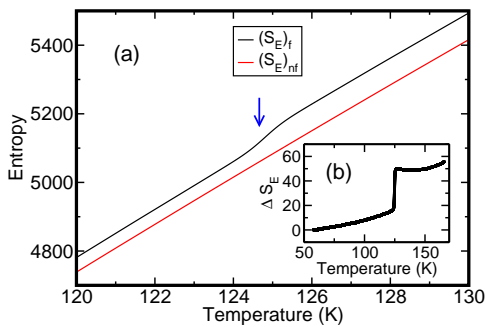}
\caption{ (a) The entropies as functions of the temperature of the
48$\times$4$\times$48 lattice with (black line) and without (red
line) frustrated spin interactions. (b) The entropy difference
between the systems with/without frustrated spin interactions. }
\label{fig:entropy}
\end{figure}

It is initially surprising that the AFM2 phase is not stable in the
small simulation lattice but survives in the large simulation
lattice. Fourier analyses of the spin structures suggest that, in
the $n$=48 lattices, many incommensurate spin components are
present, which are forbidden in the small (e.g., $n$=12) lattices.
The incommensurate spin components are generally known to arise from
the frustrated spin interactions, which suggests that the small spin
frustration might play an important role in stabilizing the AFM2
phase. We then perform the simulations using the
48$\times$4$\times$48 lattice, but set the frustrated spin
interactions to zero, i.e., $J_3$=$J_4$=$J_6$=0. The results are
shown in Fig. \ref{fig:phase-diag1}. Remarkably, the results are
very similar to those obtained for the $n$=12 lattice, and no stable
AFM2 phase is found.
We also carry out simulations by using the effective Hamiltonian and
parameters given in Ref.\cite{giovannetti11} and obtain similar results.
This result further confirms the assumption that
the weak spin frustration is essential to stabilize the AFM2 phase,
and contradicts to previous conclusions
\cite{giovannetti11,toledano11}.

To understand why the weak incommensurate spin components can
stabilize the AFM2 phase, we calculate the entropy of the system
as a function of the temperature using a multihistogram reweighting
technique\cite{ferrenberg89}. The entropies of the
48$\times$4$\times$48 lattice with or without spin frustration are
compared in Fig. \ref{fig:entropy}. The entropy of the
non-frustrated system increases smoothly as temperature increases
from 120 K to 130 K. The entropy of the $n$=12 lattices with
frustrated interactions (not shown) exhibits a similar temperature
dependence. In contrast, an obvious entropy jump occurs near
$T_1$=125 K in the 48$\times$4$\times$48 lattice with spin
frustration. This increase can be seen more clearly in Fig.
\ref{fig:entropy} (b), which depicts the difference of the system
entropies with and without spin frustrations in the temperature
range 50 -160 K. Therefore, we conclude that the incommensurate spin
excitations caused by the spin frustration greatly enhances the
entropy of the system in the AFM2 phase and stabilizes the phase.

To explore the role of the DM interaction in the phase transitions,
we tune the DM interaction strength $D$ from zero to $D_c$. In the
range of $D < D_c$, transitions from the paramagnetic-PE phase
to the AFM2-ferroelectric
phase to the AFM1-PE phase are always present. As $D$ increases, the
temperature range of the AFM2 phase also increases. When $D
>D_c$, only the transition from the paramagnetic-PE phase to
AFM2-ferroelectric phase exist,
and there is no AFM1 phase \cite{giovannetti11}. Interestingly, when
we set $D$=0, we still obtain a stable AFM2-like incommensurate
phase, but it has no net electric polarization because there is
no unique polarization axis in this case. The DM interaction that
breaks the rotational symmetry of the nearly degenerate spin
structures in the AFM2 phase and generates the polarization axis
\cite{giovannetti11}. However, in the case of a strong magnetic
anisotropy, electric polarization is also possible.

Understanding the mechanism of the multiferroicity in CuO provides
important guidance in the search for new high-T$_c$ magnetic ferroelectric
materials. We found that CuO contains two magnetic sublattices, with
strong intra-sublattice interactions and weakly frustrated
intersublattice interactions, 
which may represent one of the main reasons that the compound has a high ordering temperature.
Monte Carlo simulations suggest that the incommensurate spin excitations,
caused by the weak frustrated interactions dramatically enhance the
entropy of the multiferroic AFM2 phase, and they eventually
stabilizes the phase. The DM interaction break the magnetic
rotational symmetry, which causes lattice distortion and lead to
electric polarization in the AFM2 phase. This mechanism is distinct
from previously proposed mechanisms \cite{giovannetti11,toledano11}for CuO.
One of the fascinating features of magnetic ferroelectric
materials is their rich phase diagrams, which arises from the
competing interactions in these materials. However, few reports on
the mechanism of these phase transitions at the microscopic level
have been published.
The methods developed in this work may be
useful in elucidating the complex magnetic ferroelectric phase
transitions in general magnetic ferroelectric materials.


LH acknowledges the support of the Chinese National
Fundamental Research Program 2011CB921200, the
National Natural Science Funds for Distinguished Young Scholars,
and the Fundamental Research Funds for the Central Universities
No. WK2470000006.



\begin{thebibliography}{24}
\bibitem[{foo({\natexlab{a}})}]{footnote2}
\bibinfo{note}{The simulations are done in a 12$\times$12$\times$12 lattice,
  with magnetic interactions only.}
\bibitem[{foo({\natexlab{b}})}]{footnote1}
\bibinfo{note}{We adjust the parameters of the model Hamiltonian slightly from
  the fitted values, because first-principle calculations may not be accurate
  enough in this situation with subtle competing interactions.}
\end{thebibliography}

\end{document}